\documentclass[pra,10pt,nofootinbib, onecolumn,notitlepage, superscriptaddress]{revtex4-1}

\pdfoutput=1

\usepackage{amsmath}
\usepackage{amsfonts}
\usepackage{braket}
\usepackage{amssymb}
\usepackage{mathbbol}
\usepackage{graphicx}
\usepackage[T1]{fontenc} 
\usepackage{slashed}
\graphicspath{{.}{figures/}} 

\usepackage[colorlinks=true,linkcolor=blue, citecolor=blue, urlcolor=blue, bookmarks]{hyperref}

\newcommand{\sh}[1]{\slashed{#1}}

\begin{document}
	
	\title{Chiral transition and the chiral charge density of the hot and dense QCD matter.}
	
	\author{Chao Shi}
	\affiliation{Department of Nuclear Science and Technology, Nanjing University of Aeronautics and Astronautics, Nanjing 210016, China}
	\affiliation{Collaborative Innovation Center of Radiation Medicine of Jiangsu Higher Education Institutions, Nanjing 211106, China}
	\author{Xiao-Tao He}
        \affiliation{Department of Nuclear Science and Technology, Nanjing University of Aeronautics and Astronautics, Nanjing 210016, China}
	\author{Wen-Bao Jia}
        \affiliation{Department of Nuclear Science and Technology, Nanjing University of Aeronautics and Astronautics, Nanjing 210016, China}
	\affiliation{Collaborative Innovation Center of Radiation Medicine of Jiangsu Higher Education Institutions, Nanjing 211106, China}
	\author{Qing-Wu Wang}
        \affiliation{College of Physics, Sichuan University, Chengdu 610064, China}
	\author{Shu-Sheng Xu}
        \affiliation{College of Science, Nanjing University of Posts and Telecommunications, Nanjing 210023, China}
	\author{Hong-Shi Zong}
\affiliation{Department of Physics, Nanjing University, Nanjing 210093, China}
\affiliation{Department of Physics, Anhui Normal University, Wuhu, Anhui 241000, China}
\affiliation{Nanjing Institute of Proton Source  Technology, Nanjing 210046}

	\begin{abstract}
		
We study the chirally imbalanced hot and dense strongly interacting matter by means of the Dyson-Schwinger equations (DSEs).
The chiral phase diagram is studied in the presence of chiral chemical potential $\mu_5$. The chiral quark condensate $\langle \bar{\psi} \psi \rangle$ is obtained with the Cornwall-Jackiw-Tomboulis (CJT) effective action in concert with the Rainbow truncation. Catalysis effect of dynamical chiral symmetry breaking (DCSB) by $\mu_5$ is observed. We examine with two popular gluon models and consistency is found within the DSE approach, as well as in comparison with lattice QCD. The CEP location $(\mu_E,T_E)$ shifts toward larger $T_E$ but constant $\mu_E$ as $\mu_5$ increases. A technique is then introduced to compute the chiral charge density $n_5$ from the fully dressed quark propagator. We find the $n_5$ generally increases with temperature $T$, quark number chemical potential $\mu$ and $\mu_5$. Since the chiral magnetic effect (CME) is typically investigated with peripheral collisions, we also investigate the finite size effect on $n_5$ and find an increase in $n_5$ with smaller system size.
	\end{abstract}
	
	\maketitle

	
\section{Introduction}
\label{sec:intro}

In the heavy ion collision (HIC), a non-vanishing chiral charge $N_5$ may be induced through the Adler-Bell-Jackiw  anomaly \cite{Adler:1969gk,Bell:1969ts,Smilga:1991xa} due to topologically non-trivial gluon configurations \cite{McLerran:1990de,Shuryak:2002qz} 
\begin{align}
N_5=N_R-N_L=-\frac{g^2 N_F}{32 \pi^2}\int d^4x \epsilon _{\mu \nu \lambda \sigma} F_a^{\mu \nu} F_a^{\lambda \sigma}.
\end{align}
The $N_{R,L}$ denotes the net number of quarks (minus antiquarks) with right- or left-handed chirality, so $N_5$ is the net number of right handed quark over left handed ones. Non-vanishing $N_5$ in a strong magnetic field could result in the chiral magnetic effect (CME) \cite{Fukushima:2008xe,Kharzeev:2007jp,Copinger:2018ftr}, i.e., an electric current can be induced along the direction of the magnetic field. Consequently, there will be a charge separation within the produced fireball. Peripheral HICs provide a good testing ground for the CME as an extremely strong magnetic field $e B$ between several $m_\pi^2$ to $15\ m_\pi^2$ \cite{Skokov:2009qp, Voronyuk:2011jd, Bzdak:2011yy, Deng:2012pc,Cheng:2019qsn,Xu:2020sui} is generated by the colliding ions, in particular the spectator protons. Confirmation of the CME would reflect the local parity and charge-parity violation in quantum chromodynamics (QCD), hence is of great interest. Experimental searches have thus been actively ongoing \cite{Adamczyk:2013hsi,Abelev:2012pa,Adamczyk:2014mzf,Skokov:2016yrj}. 

To facilitate the study involving the chiral imbalance, a chiral chemical potential $\mu_5$ is introduced as conjugate to $N_5$. The associated term  $\mu_5\bar{\psi}\gamma_4\gamma_5\psi$ is then added to the Lagrangian density \cite{Fukushima:2008xe}. Technically the $N_5$ is not conserved in QCD, so the $\mu_5$ serves to mimic the chiral imbalance. It gains support from arguments that the chiral charge density $n_5$ equilibrates shortly after the collision and stay unchanged in a thermodynamical equilibrium in a longer period \cite{Ruggieri:2016asg,Ruggieri:2016lrn}. Consequently, the phase diagram of the QCD matter gets extended to a new dimension $\mu_5$, in addition to the temperature $T$ and quark number chemical potential $\mu$. However, there has been notable contradictions among different calculations, in particular concerning the chiral transition at finite temperature. The debate is over whether the pseudo-critical temperature $T_c$ (defined as the maxima of susceptibilities at finite $T$ with $\mu=0$)  increases with increasing $\mu_5$, or the opposite. The DSEs \cite{Wang:2015tia,Xu:2015vna} and lattice QCD \cite{Braguta:2015zta,Braguta:2015owi} have been giving consistent predictions that $T_c$ increases with $\mu_5$, while for NJL model the results differ by regularization schemes \cite{Ruggieri:2011xc,Yu:2015hym,Farias:2016let,Cui:2016zqp,Khunjua:2018jmn,Yang:2019lyn}. This problem is also connected with the determination of the CEP. Early model studies \cite{Chernodub:2011fr, Ruggieri:2011xc} suggest that the chiral crossover would turn into a first order phase transition at large $\mu_5$, which could be an indirect signal for the existence of CEP. However, lattice simulation finds no signal of phase transition as $\mu_5$ increases \cite{Yamamoto:2011gk, Braguta:2015zta,Braguta:2015owi}, and the DSEs found the crossover behavior persists \cite{Wang:2015tia,Xu:2015vna}.  In those work, the separable model \cite{Burden:1996nh} and the Maris-Tandy (MT) model \cite{Maris:1997tm,Maris:1999nt} were employed for gluon propagator. However, the former model is oversimplified for  purpose of computation, and the later's infrared momentum behavior contradicts today's gauge sector study \cite{Aguilar:2010gm, Boucaud:2010gr,Oliveira:2010xc,Bowman:2004jm}. So in this work we will check the consistency within DSEs by supplementing a calculation based on a more realistic gluon model, the so called Qin-Chang (QC) model. It has the correct infrared momentum behavior and also had been used extensively in hadron studies and finite temperature QCD \cite{Qin:2011dd}. However, it has never been employed in the study of finite $\mu_5$ before. Model details will be given in later sections.

The fully dressed quark propagator encodes abundant information of the QCD matter's thermodynamical properties. Among them the finite chiral charge density $n_5$ is a novel feature of the chirally imbalanced matter.  It is indispensable for the CME. Relating $n_5$ and $\mu_5$ is useful for expressing the induced electric current density as a function of the chirality density \cite{Fukushima:2010fe}. We therefore focus $n_5$ in various conditions, e.g., $T$, $\mu$, $\mu_5$ and also the system size. Note that the chemical potential $\mu$ is directly computable in the DSEs, so it doesn't pose challenge to DSE as for lattice QCD. Combined effort with DSE and lattice QCD had been carried out to locate the CEP \cite{Fischer:2010fx,Fischer:2011mz}. Meanwhile, the finite size effect could be relevant for the CME experiments since the CME is typically investigated with peripheral collisions. Note that the finite size effect on QCD phase diagram on $T-\mu$ plane had been studied with various methods \cite{Bhattacharyya:2014uxa,Tripolt:2013zfa,Braun:2011iz, Shi:2018swj,Shi:2018tsq,Abreu:2019czp,Ya-Peng:2018gkz}, but its influence on the chirally imbalanced medium was seldom studied. For a first investigation, we will consider the system in a cubic box of edge length $L$. 

This paper is organized as follows. In section \ref{sec:gap} we introduce the quark's DSE at finite $T$, $\mu$ and $\mu_5$ and its solution. In section \ref{sec:chiral} we study the chiral phase diagram in the presence of $\mu_5$. With two popular gluon propagator models, the catalysis effect of DCSB by $\mu_5$ is examined and the shift of CEP with $\mu_5$ is also shown. The $n_5$ is studied in section \ref{sec:n5}, including the finite volume effects. Finally we summarize in section \ref{sec:summary}.

\section{Quark DSE at finite chiral chemical potential}\label{sec:gap}
The Dyson-Schwinger equation of quark, namely the gap equation, at finite $T$, $\mu$ and $\mu_5$ reads \cite{Wang:2015tia}

\begin{eqnarray}
\label{eq:gapeqinf}
\hspace{-5mm} [ G(\vec{p},\tilde{\omega}_n)]^{-1}&=&[G^0(\vec{p},\tilde{\omega}_n)]^{-1}+T\sum_{l=-\infty}^\infty\int\frac{d^3q}{(2\pi)^3}
\nonumber\\
&&\hspace*{10mm}\times  \left[g^2D_{\mu \nu}(\vec{p}\!-\!\vec{q},\tilde{\omega}_n-\tilde{\omega}_l)\frac{\lambda^a}{2}\gamma_{\mu}G(\vec{q},\tilde{\omega}_l)\Gamma_{\nu}^a \right].
\end{eqnarray}
Here $G(\vec{p},\tilde{\omega}_n)$ is the fully dressed quark propagator and $G^0(\vec{p},\tilde{\omega}_n)$ is the free quark propagator. The $D_{\mu \nu}$ is the fully dressed gluon propagator (with color index contracted) and $\Gamma_{\nu}^a$ is the fully dressed quark-gluon vertex. The $\tilde{\omega}_n$ denotes $\omega_n+i \mu$, with $\omega_n$ quark's Matsubara frequency $\{\omega_n$=$(2n+1)\pi T$,  $n=0,\pm 1,\pm 2...\}$. The gluon's Matsubara frequency is $\{\Omega_n$=$2 n\pi T$, $n=0,\pm 1,\pm 2...\}$.  We've set all the renormalization constants to $1.0$ since we will use gluon models that are heavily suppressed in the ultraviolet region.  

The fully dressed quark propagator $G(\vec{p},\tilde{\omega}_n)$ can be generally decomposed as the summation of eight Dirac structures associated with coefficients of scalar functions \cite{Wang:2015tia}

\begin{subequations}\label{eq:Gdecomp}
\begin{align}
\label{eq:Ginv}
G^{-1}(\vec{p},\tilde{\omega}_n)=\sum_{i=1}^{8}T_i(\vec{p},\tilde{\omega}_n) F_i(|\vec{p}|,\tilde{\omega}_n),
\end{align}
or equivalently 
\begin{align}
\label{eq:G}
G(\vec{p},\tilde{\omega}_n)=\sum_{i=1}^{8}T_i(\vec{p},\tilde{\omega}_n) \sigma_i(|\vec{p}|,\tilde{\omega}_n).
\end{align}
\end{subequations}
The Dirac bases are $T_i \in \{i\sh{\vec{p}}, I_4, i\gamma_4 \tilde{\omega}_n, \sh{\vec{p}} \gamma_4, i \gamma_5 \sh{\vec{p}}, \gamma_5, \gamma_5  \gamma_4, i \gamma_5 \sh{\vec{p}} \gamma_4   \}$, with the last four Dirac bases brought about by the presence of chiral chemical potential $\mu_5$. The $F_i(|\vec{p}|,\tilde{\omega}_n)$'s and $\sigma_i(|\vec{p}|,\tilde{\omega}_n)$'s are dressed scalar functions. For the free quark propagator $G^0(\vec{p},\tilde{\omega}_n)$, one has $F_1=F_3=1$, $F_2=m$ and $F_7=\mu_5$, with $m$ the current quark mass. 

In \cite{Shi:2014zpa} we've studied the QCD chiral phase diagram on the $T-\mu$ plane. As a generalization to the case of nonzero $\mu_5$,  we adopt the same setup as in \cite{Shi:2014zpa}, i.e., taking the Rainbow truncation for the quark-gluon vertex
\begin{equation}
\label{eq:rainbow}
\Gamma^a_\nu(p,q)=\frac{\lambda^a}{2}\gamma_\nu,
\end{equation}
and  the fully dressed gluon propagator takes the model of
\begin{equation}
\label{eq:MT}
g^2D_{\mu\nu}(k)=\frac{4\pi^2}{\omega^6}D\textrm{e}^{-(k^2+\alpha \mu^2)/\omega^2} k^2\left(\delta_{\mu\nu}-\frac{k_\mu k_\nu}{k^2}\right).
\end{equation}
This model is based on the popular MT model that  successfully describes many hadron properties \cite{Maris:2003vk}. The additional term $e^{-\alpha \mu^2/\omega^2}$ mimics the screening effect of finite chemical potential \cite{Chen:2011my,Jiang:2013xwa}.  We remind that  the preferred parameters $\omega=0.45$ GeV, $D \omega=(0.8 \ \textrm{GeV})^3$ and $\alpha=0.6$ were used in \cite{Shi:2014zpa}, with the current quark mass set to $m=5$ MeV. Note that $\omega$ here is a parameter instead of the Matsubara frequency $\tilde{\omega}_n$.  The functions $F_i(\vec{p}, \tilde{\omega}_n)$ can then be solved: In the gap equation (\ref{eq:gapeqinf}), expand $[ G(\vec{p},\tilde{\omega}_n)]^{-1}$ with Eq.~(\ref{eq:Ginv}) and $G(\vec{q},\tilde{\omega}_l)$ with Eq.~(\ref{eq:G}), multiply both sides of Eq.~(\ref{eq:gapeqinf}) with the eight Dirac basis $T_i(\vec{p},\tilde{\omega}_n)$  and then take trace. One finally obtains eight coupled equations

\begin{subequations}\label{eq:sigtoF}
\begin{align}
F_1&=1+\frac{4}{-3\vec{p}^2}\sum \hspace{-0.5cm}\int 
\left[\sigma_1 \vec{p}\cdot\vec{q} \left(\tilde{\omega }_{ln }^2+3 \vec{k}^2\right)-2 \sigma_3 \vec{k} \cdot \vec{p}\tilde{\omega }_l \tilde{\omega }_{ln }\right]\times D_{\textrm{MT}}(\vec{k},\tilde{\omega}_{ln}) \\
F_2&=m+\frac{4}{3}\sum \hspace{-0.5cm}\int 
\left[3 \sigma_2 \left(\tilde{\omega }_{ln }^2+\vec{k}^2\right)\right] \times D_{\textrm{MT}}(\vec{k},\tilde{\omega}_{ln})\\
F_3&=1+\frac{4}{-3\tilde{\omega }_{n}^2}\sum \hspace{-0.5cm}\int
\left[\sigma_3 \tilde{\omega }_l \tilde{\omega }_n \left(3 \tilde{\omega }_{ln }^2+\vec{k}^2\right)-2 \sigma_1 \vec{k} \cdot \vec{q} \tilde{\omega }_n \tilde{\omega }_{ln }\right]\times D_{\textrm{MT}}(\vec{k},\tilde{\omega}_{ln}) \\
\label{eq:sig4toF4}
F_4&=\frac{4}{-3\vec{p}^2}\sum \hspace{-0.5cm}\int
\left[-\sigma_4 \vec{p}\cdot\vec{q} \left(\tilde{\omega }_{ln }^2+\vec{k}^2\right)\right]\times D_{\textrm{MT}}(\vec{k},\tilde{\omega}_{ln}) \\
F_5&=\frac{4}{3\vec{p}^2}\sum \hspace{-0.5cm}\int
\left[\sigma_5 \vec{p}\cdot\vec{q} \left(\tilde{\omega }_{ln }^2+3 \vec{k}^2\right)+2 i \sigma_7 \vec{k} \cdot \vec{p}\tilde{\omega }_{ln }\right]\times D_{\textrm{MT}}(\vec{k},\tilde{\omega}_{ln}) \\
\label{eq:sig6toF6}
F_6&=\frac{4}{3}\sum \hspace{-0.5cm}\int
\left[-3 \sigma_6 \left(\tilde{\omega }_{ln }^2+\vec{k}^2\right)\right]\times D_{\textrm{MT}}(\vec{k},\tilde{\omega}_{ln}) \\
F_7&=\mu_5+\frac{4}{-3}\sum \hspace{-0.5cm}\int
\left[\sigma_7 \left(-3 \tilde{\omega }_{ln }^2-\vec{k}^2\right)+2 i \sigma_5 \vec{k} \cdot \vec{q} \tilde{\omega }_{ln }\right]\times D_{\textrm{MT}}(\vec{k},\tilde{\omega}_{ln}) \\
F_8&=\frac{4}{3\vec{p}^2}\sum \hspace{-0.5cm}\int
\left[-\sigma_8 \vec{p}\cdot\vec{q} \left(\tilde{\omega }_{ln }^2+\vec{k}^2\right)\right] \times D_{\textrm{MT}}(\vec{k},\tilde{\omega}_{ln}) 
\end{align}
with 
\begin{align}
D_{\textrm{MT}}(\vec{k},\tilde{\omega}_{ln})=\frac{4\pi^2  D}{\omega^6}\textrm{e}^{-(\vec{k}^2+\tilde{\omega}_{ln}^2+\alpha \mu^2)/\omega^2}.\label{eq:D}
\end{align}
\end{subequations}
The abbreviations we used are $\tiny{\sum} \hspace{-0.35cm}\int=T\sum_{n=-\infty}^\infty \int \frac{d^3\vec{p}}{(2\pi)^3}$, $F_i=F_i(|\vec{p}|,\tilde{\omega}_n)$, $\sigma_i=\sigma_i(|\vec{q}|,\tilde{\omega}_l)$, $\vec{k}=\vec{p}-\vec{q}$, $\tilde{\omega}_{ln}=\tilde{\omega}_{l}-\tilde{\omega}_{n}$. The $\sigma_i(\vec{q}_{ l},\tilde{\omega}_l)$ can be obtained with $F_i(\vec{q}_{ l},\tilde{\omega}_l)$ via
\begin{subequations}\label{eq:Ftosig}
\begin{align}
\sigma_1&=-\frac{1}{2 | \vec{q}| } \left(\frac{F_1 | \vec{q}| -F_7}{t_1}+\frac{F_1 | \vec{q}| +F_7}{t_2}\right)\\
\sigma_2&=\frac{1}{2} \left( \frac{F_2-F_8 | \vec{q}| }{t_1}+\frac{F_8 | \vec{q}| +F_2}{t_2} \right)\\
\sigma_3&=-\frac{1}{2 \tilde{\omega }_l} \left( \frac{F_3 \tilde{\omega }_l-i F_5 | \vec{q}| }{t_1}+\frac{F_3 \tilde{\omega }_l+i F_5 | \vec{q}| }{t_2}\right)\\
\label{eq:F46tosig4}
\sigma_4&=-\frac{1}{2 | \vec{q}| } \left( \frac{F_4 | \vec{q}| -i F_6}{t_1}+\frac{F_4 | \vec{q}| +i F_6}{t_2}\right)\\
\sigma_5&=-\frac{1}{2 | \vec{q}| } \left(\frac{F_5 | \vec{q}| -i F_3 \tilde{\omega }_l}{t_1}+\frac{F_5 | \vec{q}| +i F_3 \tilde{\omega }_l}{t_2} \right)\\
\label{eq:F46tosig6}
\sigma_6&=-\frac{1}{2} \left(\frac{F_6+i F_4 | \vec{q}| }{t_1} +\frac{F_6-i F_4 | \vec{q}| }{t_2}\right)\\
\sigma_7&=-\frac{1}{2}\left( \frac{F_7-F_1 | \vec{q}| }{t_1}+\frac{F_1 | \vec{q}| +F_7}{t_2}\right)\\
\sigma_8&=\frac{1}{2 | \vec{q}| }\left(\frac{F_8 | \vec{q}| -F_2}{t_1}+\frac{F_8 | \vec{q}| +F_2}{t_2} \right)\\
t_1&=2 i F_3 F_5 \tilde{\omega }_l | \vec{q}|
   +F_3^2 \tilde{\omega }_l^2-2 F_8 F_2 | \vec{q}| -2 i F_4 F_6 | \vec{q}|
  +F_2^2-F_6^2+F_7^2 \nonumber\\
 &\hspace{20mm}+| \vec{q}| 
   \left(\left(F_1^2+F_4^2-F_5^2+F_8^2\right) | \vec{q}| -2 F_1 F_7\right)\\
t_2&=-2 i F_3 F_5 \tilde{\omega }_l | \vec{q}|
   +F_3^2 \tilde{\omega }_l^2+2 F_8 F_2 | \vec{q}| +2 i F_4 F_6 | \vec{q}|+F_2^2-F_6^2+F_7^2\nonumber \\
 &\hspace{20mm}+|\vec{q}| 
   \left(\left(F_1^2+F_4^2-F_5^2+F_8^2\right) | \vec{q}| +2 F_1 F_7\right) 
\end{align}
\end{subequations}
Eqs.~(\ref{eq:sigtoF},\ref{eq:Ftosig}) can be fully solved numerically by iteration. Meanwhile, there are two observations that could further simplify the computation. First, the $F_4$ and $F_6$ can simultaneously be zero, as can be read from Eqs.~(\ref{eq:sig4toF4},\ref{eq:sig6toF6},\ref{eq:F46tosig4},\ref{eq:F46tosig6}). In principle, there is a chance a non-vanishing solution exists, just as $F_2(\vec{p},\tilde{\omega}_n)$ associated with the quark mass function can be nonzero even in the chiral limit $m=0$ due to DCSB. But as we tested they are vanishingly small even if they are kept in the computation. Therefore the terms $F_4$ and $F_6$ can be set to zero. Another useful relation is that the scalar functions satisfy 
\begin{align}
F_i(|\vec{p}|,\tilde{\omega}_n)=F^*_i(|\vec{p}|,\tilde{\omega}_{-n-1}),
\end{align}
so the number of scalar functions with different $n$'s to compute are halved.

\section{Chiral phase diagram at finite chiral chemical potential}\label{sec:chiral}
As is well known, the DCSB and confinement are two key properties of QCD at low energy scale. The restoration of chiral symmetry and de-confinement are expected for QCD matter at finite temperature. The influence of $\mu_5$ on the chiral condensate has long been of interest \cite{Ruggieri:2011xc,Fukushima:2010fe}.
The quark condensate $\langle \bar{\psi}\psi \rangle$ is the order parameter of chiral phase transition in chiral limit, and an indicator in the presence of a small current quark mass. It can be calculated as the trace of full quark propagator
\begin{subequations}
\begin{align}
\langle \bar{\psi}\psi \rangle=T\int\frac{d^3\vec{p}}{(2\pi)^3} \sum\limits_{n=-\infty}^{\infty} \textrm{Tr}[G(\vec{p},\tilde{\omega}_n)].
\end{align}
However, this quantity suffers from ultraviolet divergence in the presence of nonzero quark mass, so people define regularized condensates as \cite{Shi:2016koj}
\begin{align}
\langle \bar{\psi}\psi \rangle_r=\langle \bar{\psi} \psi \rangle(T,\mu)-\langle \bar{\psi}_0 \psi_0 \rangle(T,\mu)
\end{align}
or that proposed in lattice calculation\cite{Bali:2011qj}
\begin{align}\label{eq:latcond}
\langle \bar{\psi}\psi \rangle_R=\langle \bar{\psi} \psi \rangle(T,\mu)-\langle \bar{\psi}_0 \psi_0 \rangle(T=0,\mu=0)
\end{align}
\end{subequations}
The $\psi_0$ indicates the free quark field. Meanwhile, a rigorous definition exists as $\langle \bar{\psi}\psi \rangle$ being the first derivative of  partition function versus current quark mass $m$, e.g.,
\begin{align}
\label{eq:cjtcond}
 \langle\bar{\psi}\psi\rangle=\frac{\partial \textrm{ln} \mathcal{Z}}{\partial m}.
\end{align}
Since the Rainbow truncation renders the  partition function calculable in the framework of CJT effective action \cite{Cornwall:1974vz,Roberts:2000aa}, i.e., 
\begin{eqnarray}
\label{eq:d}
\mathcal{P}(T,\mu,\mu_5)&=&\frac{T}{V} \textrm{ln} \mathcal{Z}=\frac{T}{V}\textrm{Tr}\biggr[Ln(G^{-1}G_0)-\frac{1}{2}(1-G_0^{-1}G)\biggr],
\end{eqnarray}
we employ  the definition of Eq.~(\ref{eq:cjtcond}) to calculate the quark condensate. This was first done in \cite{Shi:2014zpa} with $\mu_5$ not introduced. We note that the CJT effective action $\mathcal{P}(T,\mu,\mu_5)$ alone is ultra-violate divergent, but the difference of two actions with two masses, i.e., $\mathcal{P}(T,\mu,\mu_5;m+\delta m)-\mathcal{P}(T,\mu,\mu_5;m)$ is finite.
\begin{figure}[h!]
\centering 
\includegraphics[width=.49\textwidth]{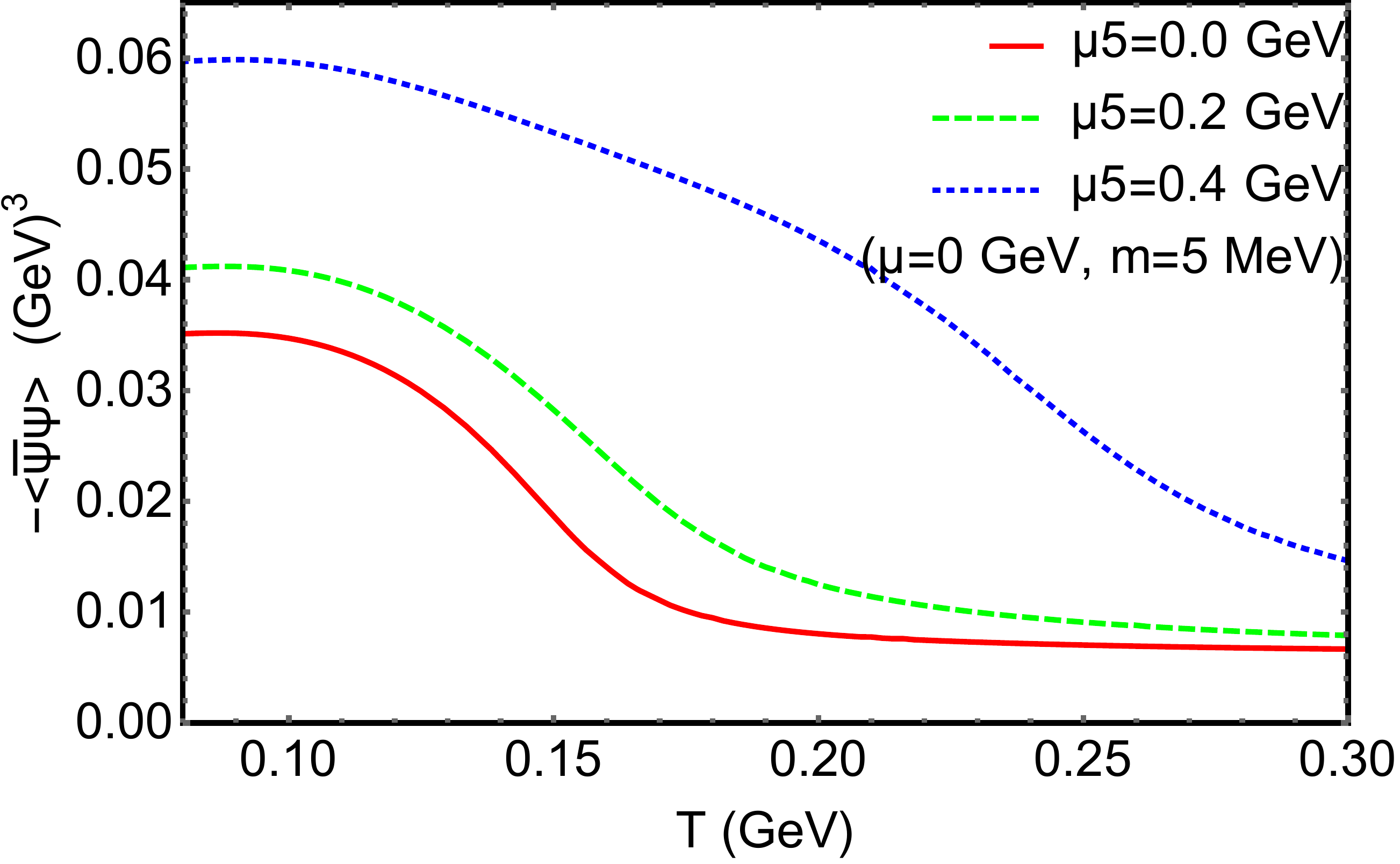}
\hfill
\includegraphics[width=.49\textwidth]{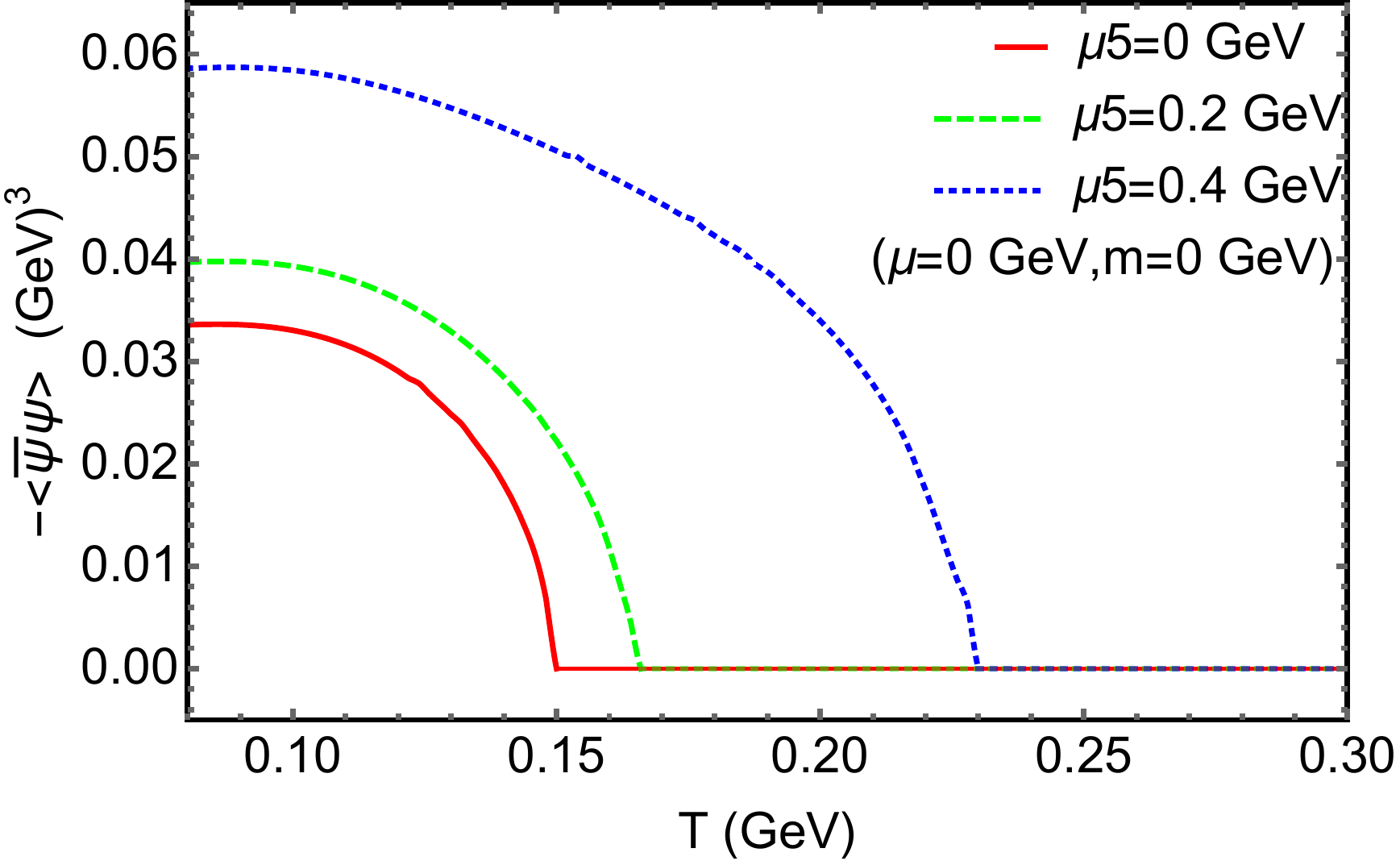}
\caption{\label{fig:condsMT} The quark condensate from MT model at finite $T$ with $\mu=0$ GeV and different  $\mu_5$'s.}
\end{figure}

\begin{figure}[h!]
\centering 
\includegraphics[width=.49\textwidth]{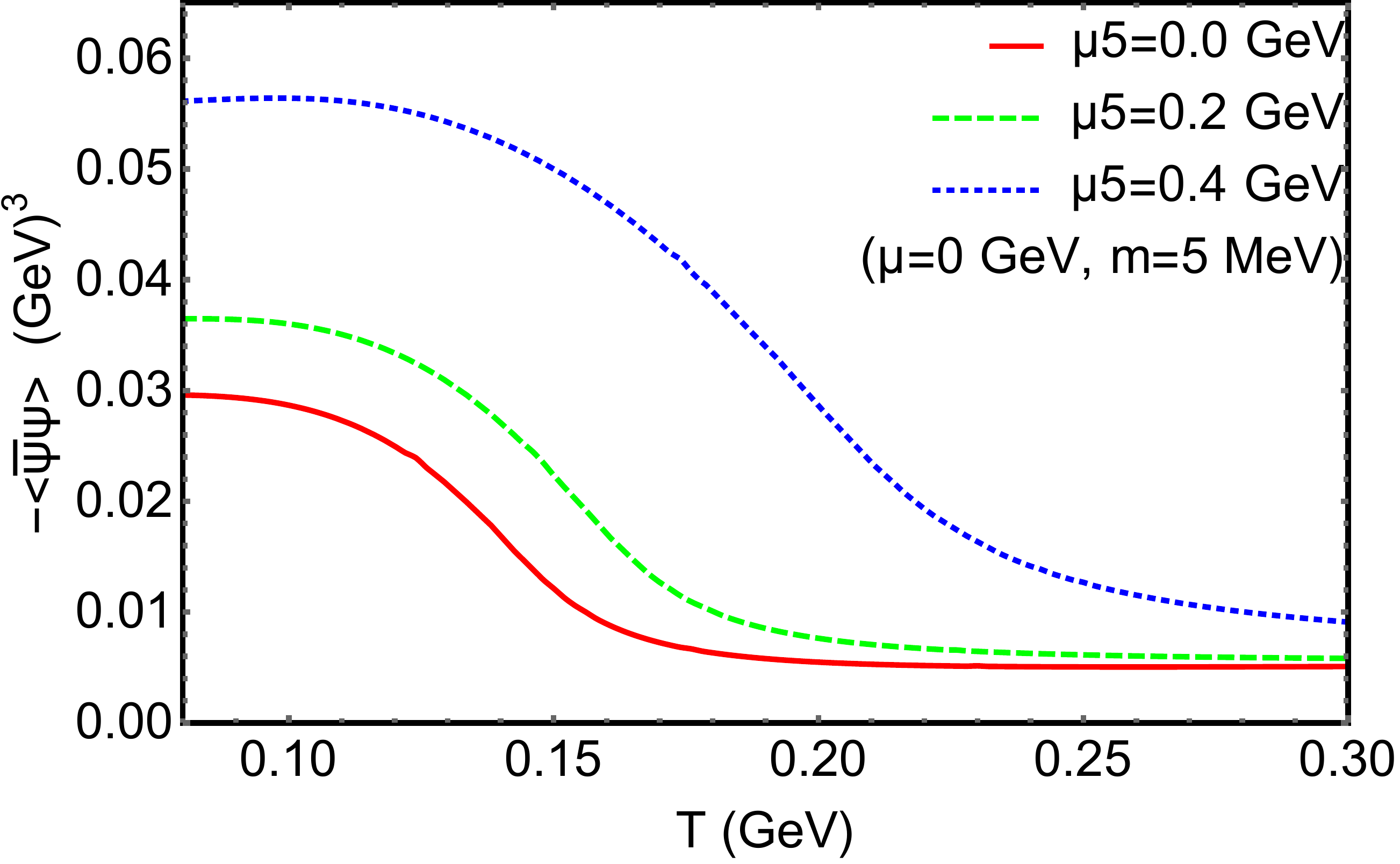}
\hfill
\includegraphics[width=.49\textwidth]{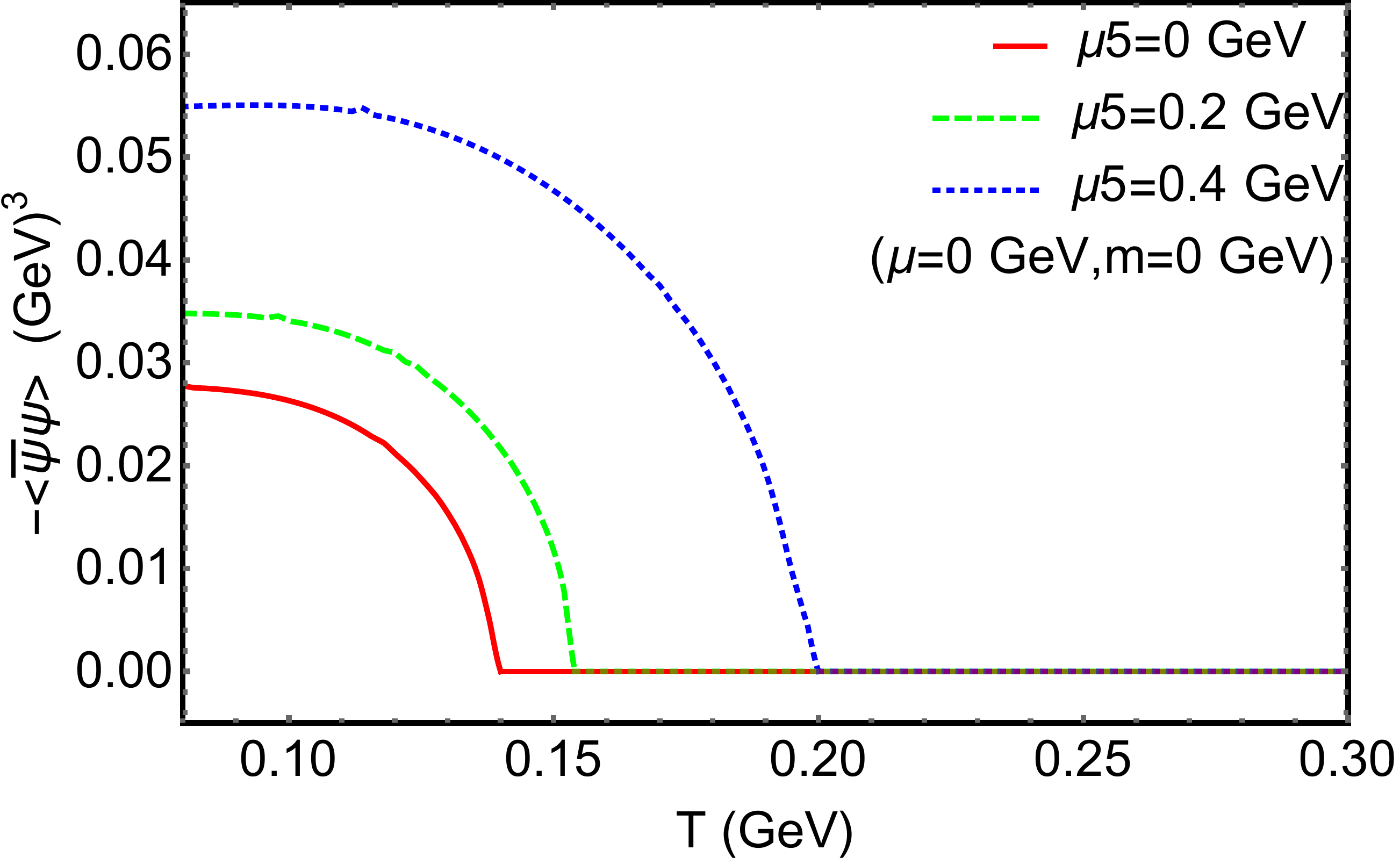}
\caption{\label{fig:condsQC}The quark condensate from QC model at finite $T$ with $\mu=0$ GeV and different  $\mu_5$'s.}
\end{figure}

The obtained $\langle\bar{\psi}\psi\rangle$ is shown in Fig.~\ref{fig:condsMT}. The left panel shows the quark condensate on the $T$ axis with different $\mu_5$'s beyond the chiral limit. An important feature is that $-\langle \bar{\psi} \psi \rangle$ rises with increasing $\mu_5$,as also found by lattice simulation \cite{Braguta:2015zta}. NJL model studies gave similar results \cite{Fukushima:2010fe,Yu:2015hym}. Meanwhile, we find the $-\langle \bar{\psi} \psi \rangle$ always exhibits crossover behavior regardless of increasing $\mu_5$, and the pseudo-transition temperature $T_c$ increases with $\mu_5$. Note that for a first order chiral phase transition the $T_c$ is unique, but for crossover it is defined as the peak location in the susceptibilities, hence definition-dependent \cite{Du:2015psa,Xu:2019ccc}. To remove the ambiguity brought by  definition of $T_c$, we supplement with the chiral limit $m=0$ case on the right panel. In this case, the chiral crossover becomes a second order phase transition. We again observe that when $\mu_5$ goes larger, the second order phase transition temperature $T_c$ increases. This is known as the catalysis effect of DCBS by $\mu_5$,  as found by lattice result as well \cite{Braguta:2015zta,Braguta:2015owi}. For other model studies as linear sigma model or NJL model, the results differ by regularization schemes \cite{Chernodub:2011fr, Yu:2015hym,Ruggieri:2011xc,Yu:2015hym,Farias:2016let,Cui:2016zqp,Khunjua:2018jmn}.

We further employ an alternative gluon propagator model, i.e., the QC model. It had never been employed in the case of a finite $\mu_5$ before. The model reads \cite{Qin:2011dd}
\begin{equation}
\label{eq:QC}
g^2D_{\mu\nu}(k)=\frac{8\pi^2}{\omega^4}D\textrm{e}^{-k^2/\omega^2} \left(\delta_{\mu\nu}-\frac{k_\mu k_\nu}{k^2}\right).
\end{equation}
Constrained by hadron properties as pion mass and decay constant, the parameters are chosen to be $D=1, \omega=0.6$, same as our early work \cite{Shi:2016koj}. This model is also widely used in hadron study within the DSE approach. Taking away the tensor part $\delta_{\mu \nu}-\frac{k_\mu k_\nu}{k^2}$, it is finite (non-vanishing) at $k^2=0$ as compared to the MT model Eq.~(\ref{eq:MT}), so it better resembles realistic gluon propagator that were produced by gauge sector study from DSE \cite{Aguilar:2010gm, Boucaud:2010gr} and lattice QCD \cite{Oliveira:2010xc,Bowman:2004jm}. The quark condensate from QC model is shown in the Fig.~\ref{fig:condsQC}. We find the QC model gives same behavior as MT model, e.g., i) the crossover behavior remains with increasing $\mu_5$ and ii) the $T_c$ rises with increasing $\mu_5$. The DSE approach therefore gives consistent result concerning the chiral crossover in the presence of chiral chemical potential.

\begin{figure}[h!]
\centering 
\includegraphics[width=.49\textwidth]{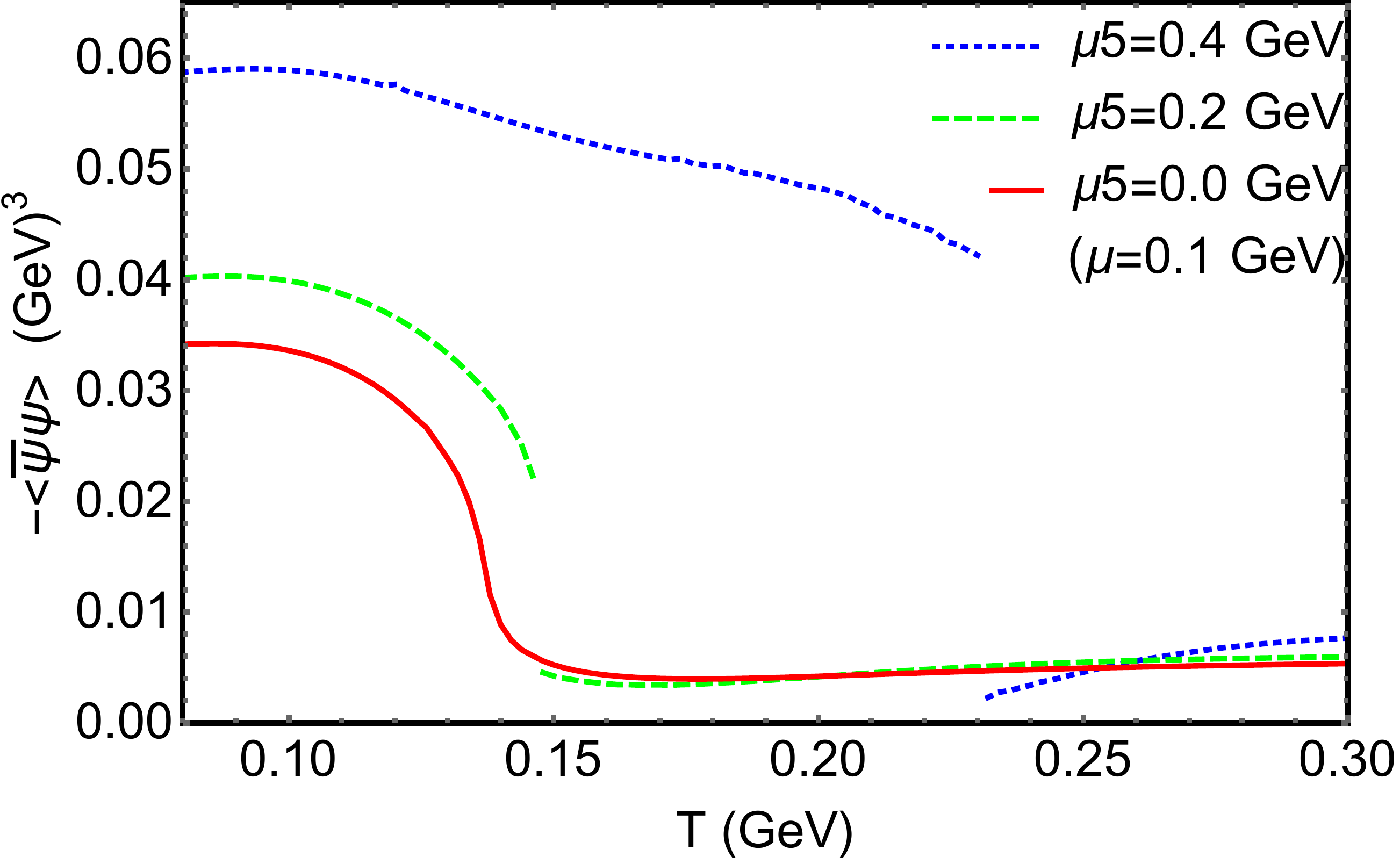}
\hfill
\includegraphics[width=.49\textwidth]{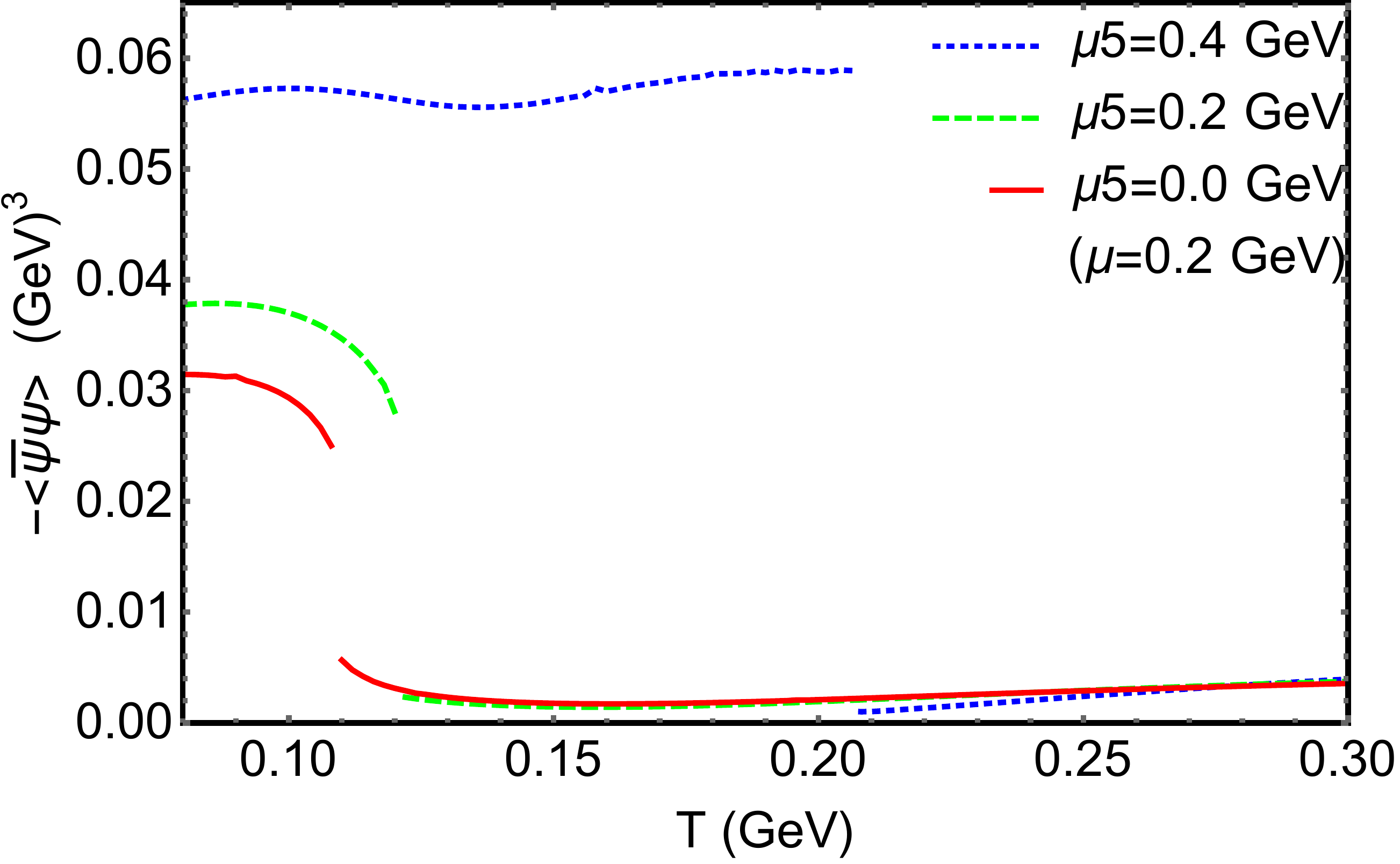}
\caption{\label{fig:condsmu}The quark condensate from MT model at finite $T$ with different $\mu$'s and $\mu_5$'s.}
\end{figure}

We also show the chiral quark condensate at finite temperature $T$ with different $\mu$'s and $\mu_5$'s in figure~\ref{fig:condsmu}. In the left panel we set $\mu=0.1$ GeV. In this case, the chiral transition is a crossover when $\mu_5=0$ GeV (red solid line), and becomes a first order phase transition when $\mu_5=0.2$ GeV (green dashed line). It remains a first order phase transition when $\mu_5$ further increases to $0.4$ GeV (blue dotted line). Therefore, the increasing $\mu_5$ changes the chiral transition at low $\mu$ from a crossover to first order phase transition. At higher $\mu$ like $\mu=0.2$ GeV in the right panel, the first order phase transition remains in the presence of finite $\mu_5$. One can observe from Fig.~\ref{fig:condsmu} that the $T_c$ generally increases with $\mu_5$ at finite $\mu$, which is similar to the $\mu=0$ case above. Note that Fig.~\ref{fig:condsmu} is plotted using the MT model. The QC model gives qualitatively same result, hence omitted. 

\begin{figure}[h!]
\centering 
\includegraphics[width=.6\textwidth]{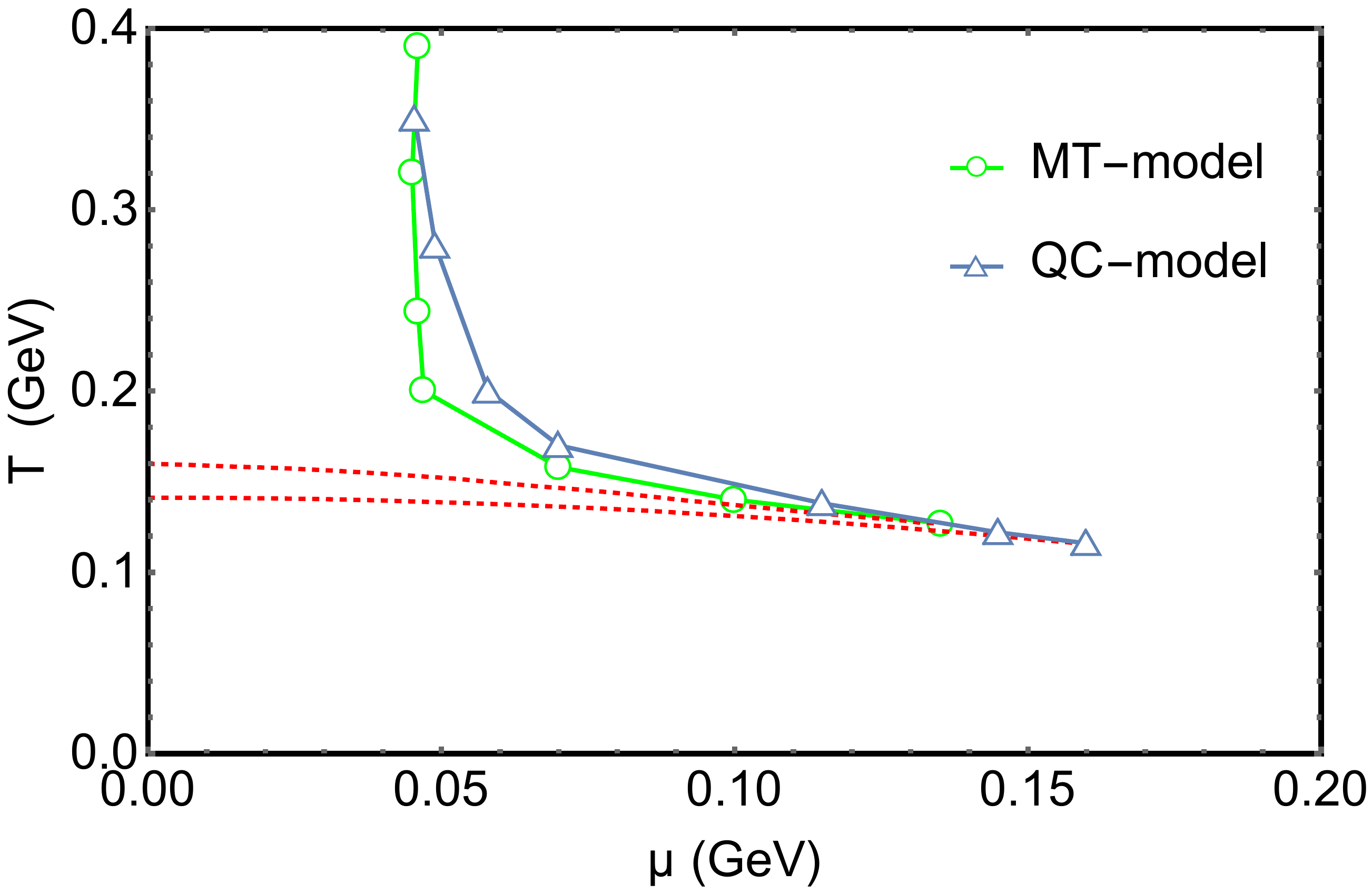}
\caption{\label{fig:CEP} The shift of CEPs in the presence of $\mu_5$. The green points (empty circle) shifting from right to left indicate the location of CEP using MT model Eq.~(\ref{eq:MT}), when $\mu_5=0, 100, 200, 300, 400, 600, 800$ MeV respectively. The blue points (empty triangle) are obtained using QC model Eq.~(\ref{eq:QC}). The red dotted lines ended with CEPs are the chiral crossover lines at zero $\mu_5$. }
\end{figure}

These findings indicate that the location of the critical end point in the $T-\mu$ plane shifts with $\mu_5$. We plot the CEPs for different $\mu_5$'s in Fig.~\ref{fig:CEP}. Results from both models are displayed. The rightmost green dot at $(\mu_T, T_E)=(0.135, 0.127)$ GeV denotes the location of CEP at $\mu_5=0$ GeV using MT model, and rightmost blue dot at $(\mu_T, T_E)=(0.16, 0.116)$ GeV denotes the one with QC model. The red dotted lines represent the corresponding crossover lines. When $\mu_5$ increases, the CEPs shift from lower-right to upper-left until the $\mu_E$'s for both models stay almost constant at $\mu_E \approx 0.05$ GeV. The CEPs therefore won't hit the temperature axis no matter how large $\mu_5$ is. Note that early model studies \cite{Fukushima:2010fe,Chernodub:2011fr,Ruggieri:2011xc} suggest that the CEP and first order phase transition would show up on the $T$ axis as $\mu_5$ increases, and confirmation from lattice QCD simulation could provide a strong evidence for the existence of CEP. However, such signal hasn't been found in existing lattice QCD \cite{Yamamoto:2011gk, Braguta:2015zta,Braguta:2015owi}. Therefore in accordance with lattice QCD at $\mu=0$, the DSE supplements with a possible trajectory of CEP at finite $\mu$, if it exists, in the presence of $\mu_5$.

\section{The chiral charge density}\label{sec:n5}

The non-vanishing finite chiral charge density $n_5$ is a novel feature of hot and dense QCD matter in the presence of $\mu_5$.  It is indispensable for the CME effect. Relating $n_5$ and $\mu_5$ is useful for to expressing the induced electric current density as a function of the chirality density \cite{Fukushima:2010fe}. The $n_5=\langle \bar{\psi}\gamma_4\gamma_5  \psi \rangle$ can be calculated with the fully dressed quark propagator
\begin{align}\label{eq:n5}
n_5=-N_c N_f T\int \frac{d^3 \vec{p}}{(2 \pi)^3}\sum_n \textrm{Tr}[\gamma_4 \gamma_5 G(\vec{p},\tilde{\omega}_n)]. 
\end{align}
However, directly computing Eq.~(\ref{eq:n5}) is very difficult: in computation one always need to truncate the Matsubara frequency $n$ to some maximum value, but Eq.~(\ref{eq:n5})  converges very slowly with $n$. Same situation happens to the computation of quark number density $n=\langle \bar{\psi}\gamma_4 \psi \rangle$ as well. In \cite{Xu:2019ccc}, a numerical technique is proposed to deal with this problem. The basic idea is to pick out the UV behavior of the integration, which converges very slowly but analytically calculable, and compute the remaining integration and summation that converges quickly. Here we take that idea and calculate $n_5$ with
\begin{align}
\label{eq:n5tech}
n_5&=\langle \bar{\psi}\gamma_4\gamma_5 \psi \rangle(T,\mu,\mu_5;m)-\langle \bar{\psi}_0\gamma_4\gamma_5 \psi_0 \rangle(T,\mu,\mu_5;m=0)+\langle \bar{\psi}_0\gamma_4\gamma_5 \psi_0 \rangle(T,\mu,\mu_5;m=0) \nonumber\\
&=N_c N_f\left(-\int \frac{d^3 \vec{p}}{(2 \pi)^3}\sum_n \textrm{Tr}\left[\gamma_4 \gamma_5 (G(\vec{p},\tilde{\omega}_n)-G_0(\vec{p},\tilde{\omega}_n;m=0))\right]\right. \nonumber \\
&\hspace{70mm}\left.+\frac{\mu_5(\pi^2 T^2+3 \mu^2+\mu_5^2)}{3\pi^2}\right).
\end{align}
Namely, we subtract the $n_5$ of free quark propagator and add it back. The summation identity 
\begin{align}
\sum\limits_{n=-\infty}^{\infty}\frac{1}{i(2n+1)\pi T-\Delta}=\frac{1}{2T}\frac{1-e^{\Delta/T}}{1+e^{\Delta/T}}
\end{align}
is useful in the derivation. In this way, the asymptotic behavior of full quark propagator $G(\vec{p},\tilde{\omega}_n)$ at large $n$ is separated out. The summation over Matsubara frequency $n$ and integration over $\vec{p}$ then converge quickly enough to validate a practical computation with precision. 

\begin{figure}[h!]
\centering 
\includegraphics[width=.6\textwidth]{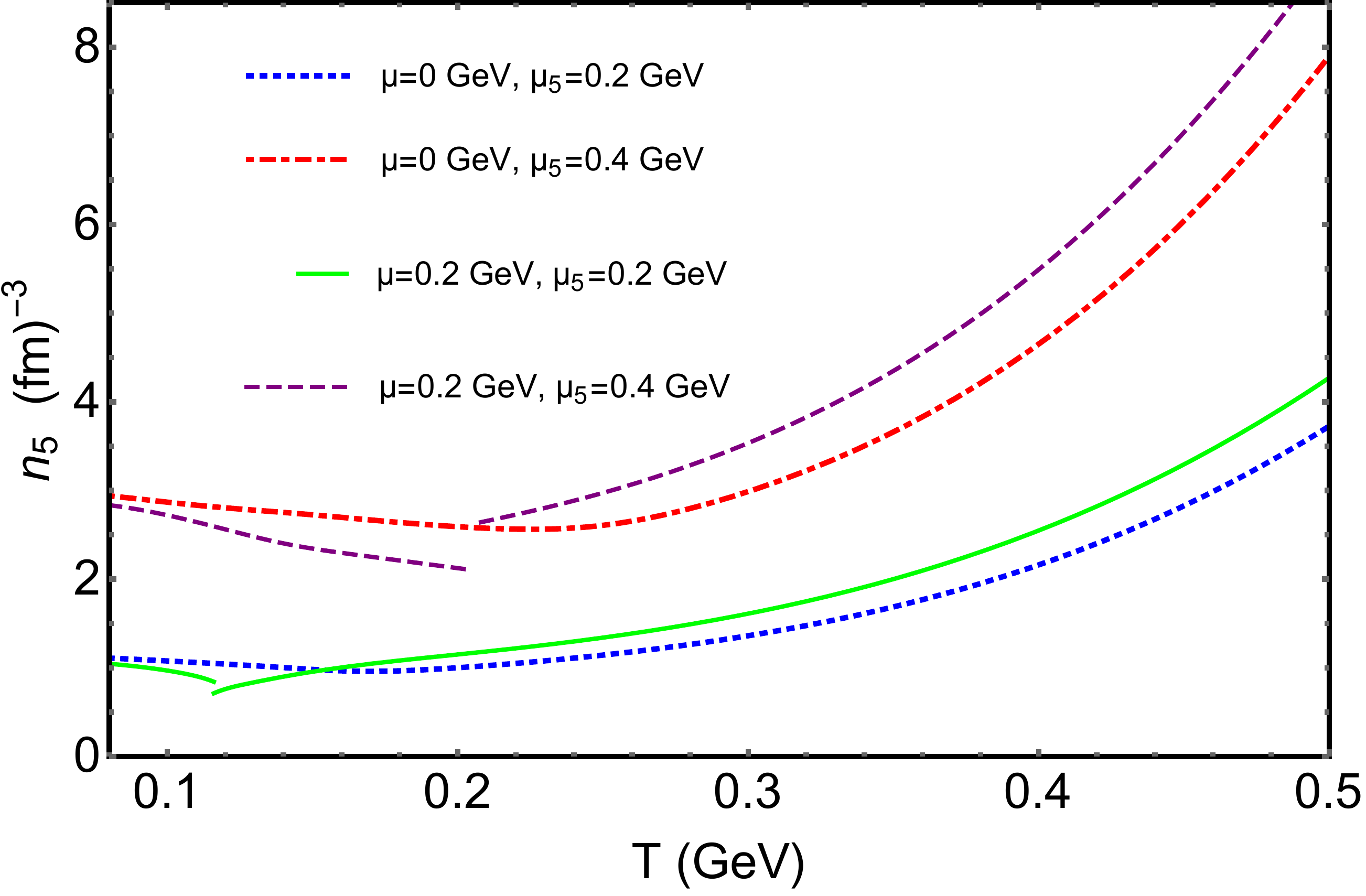}
\caption{\label{fig:n5} The chiral densities $n_5$ at finite $T$ with different $\mu$'s and $\mu_5$'s, obtained with MT model.}
\end{figure}

We then plot the chiral charge density in Fig.~\ref{fig:n5}. At $\mu_5=0$ GeV, the $n_5$ vanishes, hence not plotted. When $\mu_5=0.2$ GeV, we show the result of $\mu=0$ GeV (blue dotted) and $\mu=0.2$ GeV (green solid) respectively. One can see that $n_5$ generally increases with temperature $T$, i.e., it slightly decreases with $T$ in Nambu phase (phase with DCSB) but increases steadily in the Wigner phase (phase with chiral symmetry partially restored). The Nambu phase is irrelevant for CME, since chiral condensate couple the left-handed and right-handed quarks, and with a large chiral condensate the $n_5$ decays quickly \cite{Fukushima:2008xe}. In such case an equilibrium can not be reached. Note that de-confinement is also a necessary condition for CME, but it's beyond the scope of this work. We see in Fig.~\ref{fig:n5} that at $\mu=0$ the transition is a crossover and the $n_5$ exhibits a smooth continuous curve. At $\mu=0.2$ GeV, a first order phase transition takes place and it becomes discontinuous. Meanwhile the $n_5$ increases with $\mu_5$,  see, eg., curves from $\mu_5=0.2$ GeV to  $\mu_5=0.4$ GeV. We also observe an increase of $n_5$ with respect to quark chemical potential $\mu$ for Wigner phase in most area. Therefore, we conclude that in the Wigner phase where chiral symmetry gets partially restored, increasing $T$, $\mu_5$ and $\mu$ all result in an increase in the chiral charge density.

Another factor that can potentially influences the $n_5$ is the size of the fireball  created in HIC, namely the system volume.  Analysis shows that the volume of homogeneity before freeze-out for Au-Au and Pb-Pb collisions ranges between approximately $50\sim 250$ fm$^3$ \cite{Graef:2012sh}, and could go as low as (2 fm)$^3$ \cite{Palhares:2009tf}.  Since the CME is typically investigated with peripheral collisions, the finite size effect could be relevant. In \cite{Shi:2018swj,Shi:2018tsq} the finite volume effect on chiral phase diagram in the absence of $\mu_5$ had been studied. The finding was that decreasing the volume would weaken the DCSB, and consequently change the chiral phase diagram in the $\mu-T$ plane. Here we take the formalism and investigate the finite size effect on $n_5$. As a first estimate, we will use a rather simplified approach as in \cite{Bhattacharyya:2014uxa}, which is explained below.

At finite volume, the quark and gluon fields are constrained by certain spatial boundary condition. If we consider a system in a cubic box of size $L$, a popular and practical boundary condition is the anti-periodic boundary condition $\psi(\vec{\textbf x}=\vec{\textbf 0},\tau)=-\psi({\vec{\textbf x}=\vec{\textbf L},\tau})$  for quark fields, and $A^a_\mu(\vec{\textbf x}=\vec{\textbf 0},\tau)=A_\mu^a({\vec{\textbf x}=\vec{\textbf L},\tau})$ for gluon fields. 
In this case, the quark and gluon fields take the same boundary condition in their spatial and temporal directions.    As pointed out by authors in  \cite{Klein:2017shl}, such choice allows a permutation symmetry of the spatial and temporal directions in the effective Lagrangian, rendering temperature- and volume-independent coupling constants. One can then directly employ models that were determined at zero temperature and volume. This boundary condition therefore has been widely employed in model studies as Refs.~\cite{Fischer:2011mz,Tripolt:2013zfa,Abreu:2019czp,Shi:2018swj,Shi:2018tsq}. Mathematically, it leads to the discretization of momentum $\vec{p}$  into modes $\vec{p}_{\textbf n}=\sum_{n_i=0, \pm 1, \pm 2,...}(2 n_i+1)\pi /L\hat{e_i}$ ($\hat{e_i}$ is the Cartesian unit vectors in momentum space) for quarks, and  $\vec{p}_{\textbf n}=\sum_{n_i=0, \pm 1, \pm 2,...}2n_i\pi /L\hat{e_i}$ for gluons. So the momentum integration in Eq.~(\ref{eq:gapeqinf}) becomes
\begin{align}\label{eq:rep1}
\int \frac{dq^3}{(2 \pi)^3}(\dotsb) \rightarrow \frac{1}{L^3} \sum_{n_i=0, \pm 1, \pm 2,...} (\dotsb).
\end{align}
However, the breaking of O(3) symmetry leads to scalar functions with more variables, e.g., the $F_i(|\vec{p}|,\tilde{\omega}_n)$ in Eq.~(\ref{eq:Ginv}) now becomes $F_i(n_1,n_2,n_3,\tilde{\omega}_n)$. This calls for a lot more computing power. So for a first qualitative analyze, we use an approximation method employed by \cite{Bhattacharyya:2014uxa,Li:2017zny}, i.e.,
\begin{align}\label{eq:rep2}
\int \frac{dq^3}{(2 \pi)^3}(\dotsb) \rightarrow \int_{|\vec{q}|=\pi/L}^{|\vec{q}|=\infty} \frac{dq^3}{(2 \pi)^3}(\dotsb).
\end{align}
The right hand side (RHS) of Eq.~(\ref{eq:rep2}) intends to approximate the RHS of of Eq.~(\ref{eq:rep1}) by introducing an infrared momentum cutoff. It obviously gets more accurate at larger $L$ but less accurate with smaller $L$. However, for the purpose of a qualitative study, we find it suffices to give a correct picture in the case of chiral phase diagram. For instance, calculation with \cite{Li:2017zny} and without \cite{Shi:2018tsq} approximation give qualitatively same results concerning the chiral phase transition. 

\begin{figure}[h!]
\centering 
\includegraphics[width=.6\textwidth]{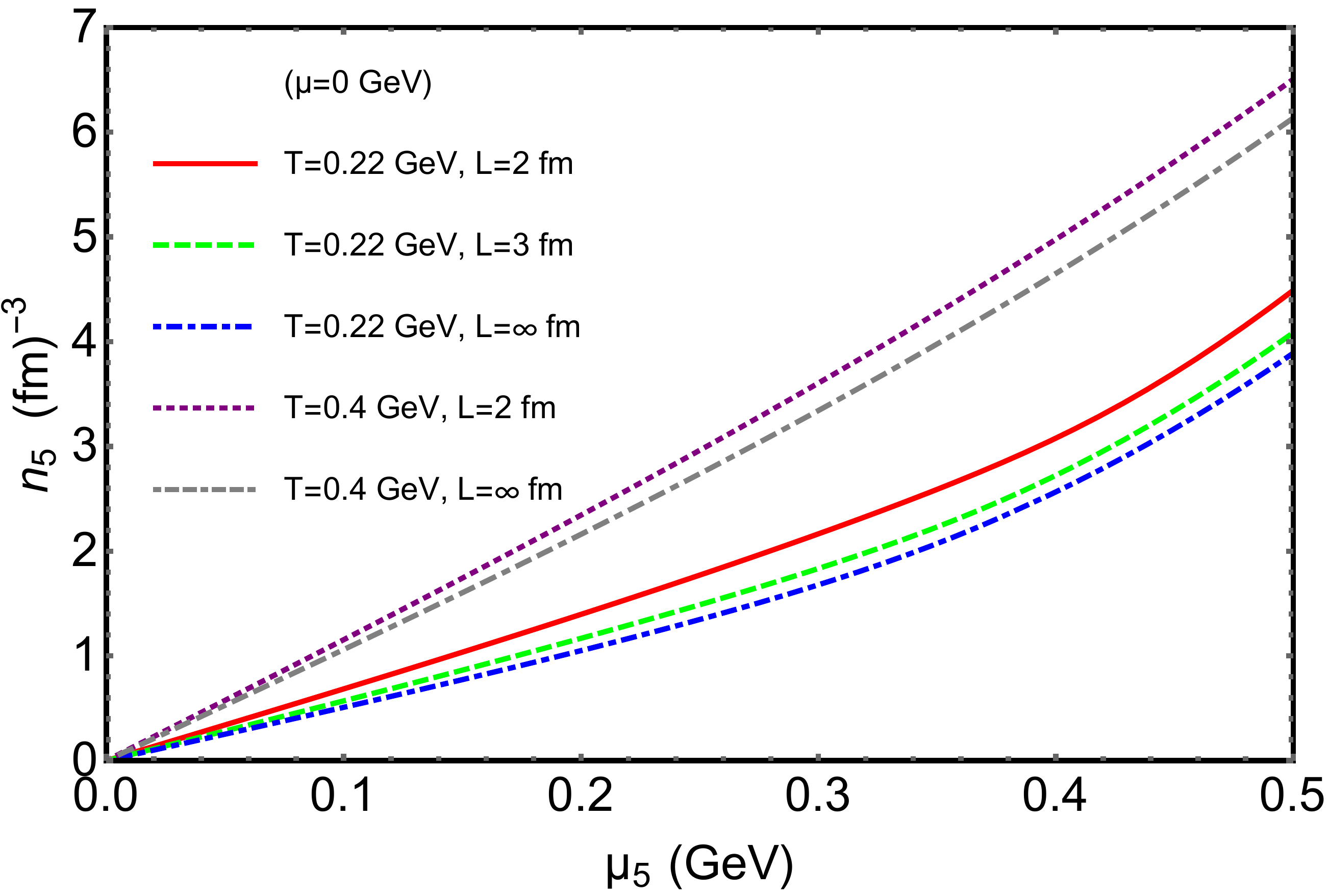}
\caption{\label{fig:n5V} The chiral density $n_5$ at finite $\mu_5$ with different system sizes, obtained by MT model.}
\end{figure}

Take the replacement Eq.~(\ref{eq:rep2}) in Eq.~(\ref{eq:gapeqinf}) and Eq.~(\ref{eq:n5}), we obtain the $n_5$ at finite volume as shown in Fig.~\ref{fig:n5V}. We remind this result is obtained with MT model at $m=5$ MeV, in correspondence with left plot in Fig.~\ref{fig:condsMT}. There are two set of curves. One is by setting $T=220$ MeV which is just above the pseudo-critical temperature $T_c \approx 150$ MeV (when $\mu=0$ and $\mu_5=0$). We see that as system size decreases, the $n_5$ increases, and the smaller the volume the quicker the increase. At $L=2$ fm, the $n_5$ increases by about 30\% for all $\mu_5$. At high temperature $T=0.4$ GeV, the finite volume effect remains but gets weakened, as can be seen from the dotted purple curve against dot-dashed gray curve. Our result therefore suggests an increase in $n_5$ when the system size decreases.

\section{Summary}\label{sec:summary}
In this paper, we study chirally imbalanced hot and dense strongly interacting matter by means of the Dyson-Schwinger equations. By solving the quark's DSE,  the fully dressed quark propagator is obtained with its complete Dirac structures, rendering thermodynamical properties calculable.

The chiral phase diagram is studied in the presence of $\mu_5$. The chiral quark condensate $\langle \bar{\psi} \psi \rangle$ is unambiguously obtained with the CJT effective action, in concert with the Rainbow truncation scheme.  Catalysis effect of DCSB by $\mu_5$ is observed. For instance, We find the $-\langle \bar{\psi} \psi \rangle$ increases with $\mu_5$, along with the increase of pseudo-critical temperature $T_c$ of the crossover at $\mu=0$. To avoid ambiguities from definition in $T_c$, we check with the chiral limit when the crossover becomes a second order phase transition. To examine the model dependence within the DSE, we supplemented a new calculation with a more realistic QC gluon propagator model, as compared to the MT model. Consistency is found within the DSE approach, as well as in comparison with lattice QCD. Since finite $\mu$ is directly calculable in the DSE, we further study the influence of $\mu_5$ on the CEP location $(\mu_E, T_E)$ in the $T-\mu$ plane. It is found that the CEP shifts toward larger $T_E$ but constant $\mu_E$ as $\mu_5$ increases. 

A technique is then developed to overcome the computational difficulty within $n_5$. We then show our results on $n_5$ with various conditions. We find the $n_5$ generally increases with the increase of $T$, $\mu$ and $\mu_5$. Since the CME is typically investigated with peripheral collisions, finite size effect is then considered based on a specific spatial boundary condition, i.e., anti-periodic for quark fields and periodic for gluon fields. We found an increase in $n_5$ with the decrease of system size. 

We remark that to mimic a more realistic condition for the HIC, the magnetic field $B$ should also be taken into account, since it influences the chiral phase diagram as well \cite{Bali:2012zg}. The finite size effect study could also be improved. For instance, based on the present boundary condition and within DSEs, a full computation using Eq.~(\ref{eq:rep1}) rather than the infrared momentum cut off scheme Eq.~(\ref{eq:rep2}) is worth studying. Meanwhile, alternative boundary conditions implementing more realistic physical constraints, such as considering a sphere or a rotating cylinder instead of cubic box, are also worth investigation \cite{Chernodub:2016kxh,Zhang:2019gva,Zhang:2020jux}. These studies could deepen our understanding of chirally imbalanced hot and dense QCD matter produced in HICs.

\acknowledgments

This work is supported in part by the National Natural Science Foundation of China (under Grants No. 11905104, No. 11475085,  No. 11535005, No. 11690030, No.11873030 and No. 11905107), the National Major state Basic Research and Development of China (Grant No. 2016YFE0129300), the starting grant of Nanjing University of Aeronautics and Astronautics (under Grant No. 1006-YAH20009), the innovation Program of Jiangsu Province, Jiangsu Province Natural Science Foundation, under grant No. BK20190721, 
Nanjing University of Posts and Telecommunications Science Foundation, under grant No. NY129032 and
Natural Science Foundation of the Jiangsu Higher Education Institutions of China 19KJB140016.

\bibliographystyle{JHEP}
\bibliography{refs}

\providecommand{\href}[2]{#2}\begingroup\raggedright\begin{thebibliography}{10}

\bibitem{Adler:1969gk}
S.~L. Adler, {\it {Axial vector vertex in spinor electrodynamics}},  {\em Phys.
  Rev.} {\bf 177} (1969) 2426--2438.

\bibitem{Bell:1969ts}
J.~Bell and R.~Jackiw, {\it {A PCAC puzzle: $\pi^0 \to \gamma \gamma$ in the
  $\sigma$ model}},  {\em Nuovo Cim. A} {\bf 60} (1969) 47--61.

\bibitem{Smilga:1991xa}
A.~V. Smilga, {\it {Anomaly mechanism at finite temperature}},  {\em Phys. Rev.
  D} {\bf 45} (1992) 1378--1394.

\bibitem{McLerran:1990de}
L.~D. McLerran, E.~Mottola, and M.~E. Shaposhnikov, {\it {Sphalerons and Axion
  Dynamics in High Temperature {QCD}}},  {\em Phys. Rev. D} {\bf 43} (1991)
  2027--2035.

\bibitem{Shuryak:2002qz}
E.~Shuryak and I.~Zahed, {\it {Prompt quark production by exploding
  sphalerons}},  {\em Phys. Rev. D} {\bf 67} (2003) 014006,
  [\href{http://arxiv.org/abs/hep-ph/0206022}{{\tt hep-ph/0206022}}].

\bibitem{Fukushima:2008xe}
K.~Fukushima, D.~E. Kharzeev, and H.~J. Warringa, {\it {The Chiral Magnetic
  Effect}},  {\em Phys. Rev. D} {\bf 78} (2008) 074033,
  [\href{http://arxiv.org/abs/0808.3382}{{\tt arXiv:0808.3382}}].

\bibitem{Kharzeev:2007jp}
D.~E. Kharzeev, L.~D. McLerran, and H.~J. Warringa, {\it {The Effects of
  topological charge change in heavy ion collisions: 'Event by event P and CP
  violation'}},  {\em Nucl. Phys. A} {\bf 803} (2008) 227--253,
  [\href{http://arxiv.org/abs/0711.0950}{{\tt arXiv:0711.0950}}].

\bibitem{Copinger:2018ftr}
P.~Copinger, K.~Fukushima, and S.~Pu, {\it {Axial Ward identity and the
  Schwinger mechanism -- Applications to the real-time chiral magnetic effect
  and condensates}},  {\em Phys. Rev. Lett.} {\bf 121} (2018), no.~26 261602,
  [\href{http://arxiv.org/abs/1807.04416}{{\tt arXiv:1807.04416}}].

\bibitem{Skokov:2009qp}
V.~Skokov, A.~Illarionov, and V.~Toneev, {\it {Estimate of the magnetic field
  strength in heavy-ion collisions}},  {\em Int. J. Mod. Phys. A} {\bf 24}
  (2009) 5925--5932, [\href{http://arxiv.org/abs/0907.1396}{{\tt
  arXiv:0907.1396}}].

\bibitem{Voronyuk:2011jd}
V.~Voronyuk, V.~Toneev, W.~Cassing, E.~Bratkovskaya, V.~Konchakovski, and
  S.~Voloshin, {\it {(Electro-)Magnetic field evolution in relativistic
  heavy-ion collisions}},  {\em Phys. Rev. C} {\bf 83} (2011) 054911,
  [\href{http://arxiv.org/abs/1103.4239}{{\tt arXiv:1103.4239}}].

\bibitem{Bzdak:2011yy}
A.~Bzdak and V.~Skokov, {\it {Event-by-event fluctuations of magnetic and
  electric fields in heavy ion collisions}},  {\em Phys. Lett. B} {\bf 710}
  (2012) 171--174, [\href{http://arxiv.org/abs/1111.1949}{{\tt
  arXiv:1111.1949}}].

\bibitem{Deng:2012pc}
W.-T. Deng and X.-G. Huang, {\it {Event-by-event generation of electromagnetic
  fields in heavy-ion collisions}},  {\em Phys. Rev. C} {\bf 85} (2012) 044907,
  [\href{http://arxiv.org/abs/1201.5108}{{\tt arXiv:1201.5108}}].

\bibitem{Cheng:2019qsn}
Y.-L. Cheng, S.~Zhang, Y.-G. Ma, J.-H. Chen, and C.~Zhong, {\it
  {Electromagnetic field from asymmetric to symmetric heavy-ion collisions at
  200 GeV/c}},  {\em Phys. Rev. C} {\bf 99} (2019), no.~5 054906,
  [\href{http://arxiv.org/abs/1909.03160}{{\tt arXiv:1909.03160}}].

\bibitem{Xu:2020sui}
K.~Xu, S.~Shi, H.~Zhang, D.~Hou, J.~Liao, and M.~Huang, {\it {Extracting the
  magnitude of magnetic field at freeze-out in heavy-ion collisions}},
  \href{http://arxiv.org/abs/2004.05362}{{\tt arXiv:2004.05362}}.

\bibitem{Adamczyk:2013hsi}
{\bf STAR} Collaboration, L.~Adamczyk et~al., {\it {Fluctuations of charge
  separation perpendicular to the event plane and local parity violation in
  $\sqrt{s_{NN}}=200$ GeV Au+Au collisions at the BNL Relativistic Heavy Ion
  Collider}},  {\em Phys. Rev. C} {\bf 88} (2013), no.~6 064911,
  [\href{http://arxiv.org/abs/1302.3802}{{\tt arXiv:1302.3802}}].

\bibitem{Abelev:2012pa}
{\bf ALICE} Collaboration, B.~Abelev et~al., {\it {Charge separation relative
  to the reaction plane in Pb-Pb collisions at $\sqrt{s_{NN}}= 2.76$ TeV}},
  {\em Phys. Rev. Lett.} {\bf 110} (2013), no.~1 012301,
  [\href{http://arxiv.org/abs/1207.0900}{{\tt arXiv:1207.0900}}].

\bibitem{Adamczyk:2014mzf}
{\bf STAR} Collaboration, L.~Adamczyk et~al., {\it {Beam-energy dependence of
  charge separation along the magnetic field in Au+Au collisions at RHIC}},
  {\em Phys. Rev. Lett.} {\bf 113} (2014) 052302,
  [\href{http://arxiv.org/abs/1404.1433}{{\tt arXiv:1404.1433}}].

\bibitem{Skokov:2016yrj}
V.~Koch, S.~Schlichting, V.~Skokov, P.~Sorensen, J.~Thomas, S.~Voloshin,
  G.~Wang, and H.-U. Yee, {\it {Status of the chiral magnetic effect and
  collisions of isobars}},  {\em Chin. Phys. C} {\bf 41} (2017), no.~7 072001,
  [\href{http://arxiv.org/abs/1608.00982}{{\tt arXiv:1608.00982}}].

\bibitem{Ruggieri:2016asg}
M.~Ruggieri, G.~Peng, and M.~Chernodub, {\it {Chiral Relaxation Time at the
  Crossover of Quantum Chromodynamics}},  {\em Phys. Rev. D} {\bf 94} (2016),
  no.~5 054011, [\href{http://arxiv.org/abs/1606.03287}{{\tt
  arXiv:1606.03287}}].

\bibitem{Ruggieri:2016lrn}
M.~Ruggieri and G.~Peng, {\it {Quark matter in a parallel electric and magnetic
  field background: Chiral phase transition and equilibration of chiral
  density}},  {\em Phys. Rev. D} {\bf 93} (2016), no.~9 094021,
  [\href{http://arxiv.org/abs/1602.08994}{{\tt arXiv:1602.08994}}].

\bibitem{Wang:2015tia}
B.~Wang, Y.-L. Wang, Z.-F. Cui, and H.-S. Zong, {\it {Effect of the chiral
  chemical potential on the position of the critical endpoint}},  {\em Phys.
  Rev. D} {\bf 91} (2015), no.~3 034017.

\bibitem{Xu:2015vna}
S.-S. Xu, Z.-F. Cui, B.~Wang, Y.-M. Shi, Y.-C. Yang, and H.-S. Zong, {\it
  {Chiral phase transition with a chiral chemical potential in the framework of
  Dyson-Schwinger equations}},  {\em Phys. Rev. D} {\bf 91} (2015), no.~5
  056003, [\href{http://arxiv.org/abs/1505.00316}{{\tt arXiv:1505.00316}}].

\bibitem{Braguta:2015zta}
V.~Braguta, V.~Goy, E.~M. Ilgenfritz, A.~Y. Kotov, A.~Molochkov,
  M.~Muller-Preussker, and B.~Petersson, {\it {Two-Color QCD with Non-zero
  Chiral Chemical Potential}},  {\em JHEP} {\bf 06} (2015) 094,
  [\href{http://arxiv.org/abs/1503.06670}{{\tt arXiv:1503.06670}}].

\bibitem{Braguta:2015owi}
V.~Braguta, E.~Ilgenfritz, A.~Y. Kotov, B.~Petersson, and S.~Skinderev, {\it
  {Study of QCD Phase Diagram with Non-Zero Chiral Chemical Potential}},  {\em
  Phys. Rev. D} {\bf 93} (2016), no.~3 034509,
  [\href{http://arxiv.org/abs/1512.05873}{{\tt arXiv:1512.05873}}].

\bibitem{Ruggieri:2011xc}
M.~Ruggieri, {\it {The Critical End Point of Quantum Chromodynamics Detected by
  Chirally Imbalanced Quark Matter}},  {\em Phys. Rev. D} {\bf 84} (2011)
  014011, [\href{http://arxiv.org/abs/1103.6186}{{\tt arXiv:1103.6186}}].

\bibitem{Yu:2015hym}
L.~Yu, H.~Liu, and M.~Huang, {\it {Effect of the chiral chemical potential on
  the chiral phase transition in the NJL model with different regularization
  schemes}},  {\em Phys. Rev. D} {\bf 94} (2016), no.~1 014026,
  [\href{http://arxiv.org/abs/1511.03073}{{\tt arXiv:1511.03073}}].

\bibitem{Farias:2016let}
R.~Farias, D.~C. Duarte, G.~Krein, and R.~O. Ramos, {\it {Thermodynamics of
  quark matter with a chiral imbalance}},  {\em Phys. Rev. D} {\bf 94} (2016),
  no.~7 074011, [\href{http://arxiv.org/abs/1604.04518}{{\tt
  arXiv:1604.04518}}].

\bibitem{Cui:2016zqp}
Z.-F. Cui, I.~C. Cloet, Y.~Lu, C.~D. Roberts, S.~M. Schmidt, S.-S. Xu, and
  H.-S. Zong, {\it {Critical endpoint in the presence of a chiral chemical
  potential}},  {\em Phys. Rev. D} {\bf 94} (2016) 071503,
  [\href{http://arxiv.org/abs/1604.08454}{{\tt arXiv:1604.08454}}].

\bibitem{Khunjua:2018jmn}
T.~Khunjua, K.~Klimenko, and R.~Zhokhov, {\it {Chiral imbalanced hot and dense
  quark matter: NJL analysis at the physical point and comparison with lattice
  QCD}},  {\em Eur. Phys. J. C} {\bf 79} (2019), no.~2 151,
  [\href{http://arxiv.org/abs/1812.00772}{{\tt arXiv:1812.00772}}].

\bibitem{Yang:2019lyn}
L.-K. Yang, X.~Luo, and H.-S. Zong, {\it {QCD phase diagram in chiral imbalance
  with self-consistent mean field approximation}},  {\em Phys. Rev. D} {\bf
  100} (2019), no.~9 094012, [\href{http://arxiv.org/abs/1910.13185}{{\tt
  arXiv:1910.13185}}].

\bibitem{Chernodub:2011fr}
M.~Chernodub and A.~Nedelin, {\it {Phase diagram of chirally imbalanced QCD
  matter}},  {\em Phys. Rev. D} {\bf 83} (2011) 105008,
  [\href{http://arxiv.org/abs/1102.0188}{{\tt arXiv:1102.0188}}].

\bibitem{Yamamoto:2011gk}
A.~Yamamoto, {\it {Chiral magnetic effect in lattice QCD with a chiral chemical
  potential}},  {\em Phys. Rev. Lett.} {\bf 107} (2011) 031601,
  [\href{http://arxiv.org/abs/1105.0385}{{\tt arXiv:1105.0385}}].

\bibitem{Burden:1996nh}
C.~Burden, L.~Qian, C.~D. Roberts, P.~Tandy, and M.~J. Thomson, {\it {Ground
  state spectrum of light quark mesons}},  {\em Phys. Rev. C} {\bf 55} (1997)
  2649--2664, [\href{http://arxiv.org/abs/nucl-th/9605027}{{\tt
  nucl-th/9605027}}].

\bibitem{Maris:1997tm}
P.~Maris and C.~D. Roberts, {\it {Pi- and K meson Bethe-Salpeter amplitudes}},
  {\em Phys. Rev. C} {\bf 56} (1997) 3369--3383,
  [\href{http://arxiv.org/abs/nucl-th/9708029}{{\tt nucl-th/9708029}}].

\bibitem{Maris:1999nt}
P.~Maris and P.~C. Tandy, {\it {Bethe-Salpeter study of vector meson masses and
  decay constants}},  {\em Phys. Rev. C} {\bf 60} (1999) 055214,
  [\href{http://arxiv.org/abs/nucl-th/9905056}{{\tt nucl-th/9905056}}].

\bibitem{Aguilar:2010gm}
A.~Aguilar, D.~Binosi, and J.~Papavassiliou, {\it {QCD effective charges from
  lattice data}},  {\em JHEP} {\bf 07} (2010) 002,
  [\href{http://arxiv.org/abs/1004.1105}{{\tt arXiv:1004.1105}}].

\bibitem{Boucaud:2010gr}
P.~Boucaud, M.~Gomez, J.~Leroy, A.~Le~Yaouanc, J.~Micheli, O.~Pene, and
  J.~Rodriguez-Quintero, {\it {The low-momentum ghost dressing function and the
  gluon mass}},  {\em Phys. Rev. D} {\bf 82} (2010) 054007,
  [\href{http://arxiv.org/abs/1004.4135}{{\tt arXiv:1004.4135}}].

\bibitem{Oliveira:2010xc}
O.~Oliveira and P.~Bicudo, {\it {Running Gluon Mass from Landau Gauge Lattice
  QCD Propagator}},  {\em J. Phys. G} {\bf 38} (2011) 045003,
  [\href{http://arxiv.org/abs/1002.4151}{{\tt arXiv:1002.4151}}].

\bibitem{Bowman:2004jm}
P.~O. Bowman, U.~M. Heller, D.~B. Leinweber, M.~B. Parappilly, and A.~G.
  Williams, {\it {Unquenched gluon propagator in Landau gauge}},  {\em Phys.
  Rev. D} {\bf 70} (2004) 034509,
  [\href{http://arxiv.org/abs/hep-lat/0402032}{{\tt hep-lat/0402032}}].

\bibitem{Qin:2011dd}
S.-x. Qin, L.~Chang, Y.-x. Liu, C.~D. Roberts, and D.~J. Wilson, {\it
  {Interaction model for the gap equation}},  {\em Phys. Rev. C} {\bf 84}
  (2011) 042202, [\href{http://arxiv.org/abs/1108.0603}{{\tt
  arXiv:1108.0603}}].

\bibitem{Fukushima:2010fe}
K.~Fukushima, M.~Ruggieri, and R.~Gatto, {\it {Chiral magnetic effect in the
  PNJL model}},  {\em Phys. Rev. D} {\bf 81} (2010) 114031,
  [\href{http://arxiv.org/abs/1003.0047}{{\tt arXiv:1003.0047}}].

\bibitem{Fischer:2010fx}
C.~S. Fischer, A.~Maas, and J.~A. Muller, {\it {Chiral and deconfinement
  transition from correlation functions: SU(2) vs. SU(3)}},  {\em Eur. Phys. J.
  C} {\bf 68} (2010) 165--181, [\href{http://arxiv.org/abs/1003.1960}{{\tt
  arXiv:1003.1960}}].

\bibitem{Fischer:2011mz}
C.~S. Fischer, J.~Luecker, and J.~A. Mueller, {\it {Chiral and deconfinement
  phase transitions of two-flavour QCD at finite temperature and chemical
  potential}},  {\em Phys. Lett. B} {\bf 702} (2011) 438--441,
  [\href{http://arxiv.org/abs/1104.1564}{{\tt arXiv:1104.1564}}].

\bibitem{Bhattacharyya:2014uxa}
A.~Bhattacharyya, R.~Ray, and S.~Sur, {\it {Fluctuation of strongly interacting
  matter in the Polyakov--Nambu--Jona-Lasinio model in a finite volume}},  {\em
  Phys. Rev. D} {\bf 91} (2015), no.~5 051501,
  [\href{http://arxiv.org/abs/1412.8316}{{\tt arXiv:1412.8316}}].

\bibitem{Tripolt:2013zfa}
R.-A. Tripolt, J.~Braun, B.~Klein, and B.-J. Schaefer, {\it {Effect of
  fluctuations on the QCD critical point in a finite volume}},  {\em Phys. Rev.
  D} {\bf 90} (2014), no.~5 054012, [\href{http://arxiv.org/abs/1308.0164}{{\tt
  arXiv:1308.0164}}].

\bibitem{Braun:2011iz}
J.~Braun, B.~Klein, and B.-J. Schaefer, {\it {On the Phase Structure of QCD in
  a Finite Volume}},  {\em Phys. Lett. B} {\bf 713} (2012) 216--223,
  [\href{http://arxiv.org/abs/1110.0849}{{\tt arXiv:1110.0849}}].

\bibitem{Shi:2018swj}
C.~Shi, W.~Jia, A.~Sun, L.~Zhang, and H.-S. Zong, {\it {Chiral crossover
  transition in a finite volume}},  {\em Chin. Phys. C} {\bf 42} (2018), no.~2
  023101.

\bibitem{Shi:2018tsq}
C.~Shi, Y.~Xia, W.~Jia, and H.~Zong, {\it {Chiral phase diagram of strongly
  interacting matter at finite volume}},  {\em Sci. China Phys. Mech. Astron.}
  {\bf 61} (2018), no.~8 082021.

\bibitem{Abreu:2019czp}
L.~Abreu, E.~B. Corrêa, C.~A. Linhares, and A.~P. Malbouisson, {\it
  {Finite-volume and magnetic effects on the phase structure of the
  three-flavor Nambu--Jona-Lasinio model}},  {\em Phys. Rev. D} {\bf 99}
  (2019), no.~7 076001, [\href{http://arxiv.org/abs/1903.09249}{{\tt
  arXiv:1903.09249}}].

\bibitem{Ya-Peng:2018gkz}
Y.-P. Zhao, P.-L. Yin, Z.-H. Yu, and H.-S. Zong, {\it {Finite volume effects on
  chiral phase transition and pseudoscalar mesons properties from the
  Polyakov-Nambu-Jona-Lasinio model}},  {\em Nucl. Phys.} {\bf B952} (2020)
  114919, [\href{http://arxiv.org/abs/1812.09665}{{\tt arXiv:1812.09665}}].

\bibitem{Shi:2014zpa}
C.~Shi, Y.-L. Wang, Y.~Jiang, Z.-F. Cui, and H.-S. Zong, {\it {Locate QCD
  Critical End Point in a Continuum Model Study}},  {\em JHEP} {\bf 07} (2014)
  014, [\href{http://arxiv.org/abs/1403.3797}{{\tt arXiv:1403.3797}}].

\bibitem{Maris:2003vk}
P.~Maris and C.~D. Roberts, {\it {Dyson-Schwinger equations: A Tool for hadron
  physics}},  {\em Int. J. Mod. Phys. E} {\bf 12} (2003) 297--365,
  [\href{http://arxiv.org/abs/nucl-th/0301049}{{\tt nucl-th/0301049}}].

\bibitem{Chen:2011my}
H.~Chen, M.~Baldo, G.~Burgio, and H.-J. Schulze, {\it {Hybrid stars with the
  Dyson-Schwinger quark model}},  {\em Phys. Rev. D} {\bf 84} (2011) 105023,
  [\href{http://arxiv.org/abs/1107.2497}{{\tt arXiv:1107.2497}}].

\bibitem{Jiang:2013xwa}
Y.~Jiang, H.~Chen, W.-M. Sun, and H.-S. Zong, {\it {Chiral phase transition of
  QCD at finite chemical potential}},  {\em JHEP} {\bf 04} (2013) 014.

\bibitem{Shi:2016koj}
C.~Shi, Y.-L. Du, S.-S. Xu, X.-J. Liu, and H.-S. Zong, {\it {Continuum study of
  the QCD phase diagram through an OPE-modified gluon propagator}},  {\em Phys.
  Rev. D} {\bf 93} (2016), no.~3 036006,
  [\href{http://arxiv.org/abs/1602.00062}{{\tt arXiv:1602.00062}}].

\bibitem{Bali:2011qj}
G.~Bali, F.~Bruckmann, G.~Endrodi, Z.~Fodor, S.~Katz, S.~Krieg, A.~Schafer, and
  K.~Szabo, {\it {The QCD phase diagram for external magnetic fields}},  {\em
  JHEP} {\bf 02} (2012) 044, [\href{http://arxiv.org/abs/1111.4956}{{\tt
  arXiv:1111.4956}}].

\bibitem{Cornwall:1974vz}
J.~M. Cornwall, R.~Jackiw, and E.~Tomboulis, {\it {Effective Action for
  Composite Operators}},  {\em Phys. Rev. D} {\bf 10} (1974) 2428--2445.

\bibitem{Roberts:2000aa}
C.~D. Roberts and S.~M. Schmidt, {\it {Dyson-Schwinger equations: Density,
  temperature and continuum strong QCD}},  {\em Prog. Part. Nucl. Phys.} {\bf
  45} (2000) S1--S103, [\href{http://arxiv.org/abs/nucl-th/0005064}{{\tt
  nucl-th/0005064}}].

\bibitem{Du:2015psa}
Y.-L. Du, Y.~Lu, S.-S. Xu, Z.-F. Cui, C.~Shi, and H.-S. Zong, {\it
  {Susceptibilities and critical exponents within the Nambu--Jona-Lasinio
  model}},  {\em Int. J. Mod. Phys. A} {\bf 30} (2015), no.~34 1550199,
  [\href{http://arxiv.org/abs/1506.04368}{{\tt arXiv:1506.04368}}].

\bibitem{Xu:2019ccc}
S.-S. Xu, P.-L. Yin, and H.-S. Zong, {\it {Susceptibilities and the critical
  band of crossover region in the QCD phase diagram}},  {\em Eur. Phys. J. C}
  {\bf 79} (2019), no.~5 399.

\bibitem{Graef:2012sh}
G.~Graef, M.~Bleicher, and Q.~Li, {\it {Examination of scaling of
  Hanbury-Brown--Twiss radii with charged particle multiplicity}},  {\em Phys.
  Rev. C} {\bf 85} (2012) 044901, [\href{http://arxiv.org/abs/1203.4071}{{\tt
  arXiv:1203.4071}}].

\bibitem{Palhares:2009tf}
L.~Palhares, E.~Fraga, and T.~Kodama, {\it {Chiral transition in a finite
  system and possible use of finite size scaling in relativistic heavy ion
  collisions}},  {\em J. Phys. G} {\bf 38} (2011) 085101,
  [\href{http://arxiv.org/abs/0904.4830}{{\tt arXiv:0904.4830}}].

\bibitem{Klein:2017shl}
B.~Klein, {\it {Modeling Finite-Volume Effects and Chiral Symmetry Breaking in
  Two-Flavor QCD Thermodynamics}},  {\em Phys. Rept.} {\bf 707-708} (2017)
  1--51, [\href{http://arxiv.org/abs/1710.05357}{{\tt arXiv:1710.05357}}].

\bibitem{Li:2017zny}
B.-L. Li, Z.-F. Cui, B.-W. Zhou, S.~An, L.-P. Zhang, and H.-S. Zong, {\it
  {Finite volume effects on the chiral phase transition from Dyson--Schwinger
  equations of QCD}},  {\em Nucl. Phys. B} {\bf 938} (2019) 298--306,
  [\href{http://arxiv.org/abs/1711.04914}{{\tt arXiv:1711.04914}}].

\bibitem{Bali:2012zg}
G.~Bali, F.~Bruckmann, G.~Endrodi, Z.~Fodor, S.~Katz, and A.~Schafer, {\it {QCD
  quark condensate in external magnetic fields}},  {\em Phys. Rev. D} {\bf 86}
  (2012) 071502, [\href{http://arxiv.org/abs/1206.4205}{{\tt
  arXiv:1206.4205}}].

\bibitem{Chernodub:2016kxh}
M.~Chernodub and S.~Gongyo, {\it {Interacting fermions in rotation: chiral
  symmetry restoration, moment of inertia and thermodynamics}},  {\em JHEP}
  {\bf 01} (2017) 136, [\href{http://arxiv.org/abs/1611.02598}{{\tt
  arXiv:1611.02598}}].

\bibitem{Zhang:2019gva}
Z.~Zhang, C.~Shi, and H.-S. Zong, {\it {Nambu-Jona-Lasinio model in a sphere}},
   {\em Phys. Rev. D} {\bf 101} (2020), no.~4 043006,
  [\href{http://arxiv.org/abs/1908.08671}{{\tt arXiv:1908.08671}}].

\bibitem{Zhang:2020jux}
Z.~Zhang, C.~Shi, X.~Luo, and H.-S. Zong, {\it {Chiral phase transition in a
  rotating sphere}},  {\em Phys.Rev.D} (in press),
  [\href{http://arxiv.org/abs/2003.03765}{{\tt arXiv:2003.03765}}].

\end{thebibliography}\endgroup

\end{document}